\documentclass[aps,prd,twocolumn,groupedaddress,showpacs,nofootinbib,amssymb]{revtex4-1}
\usepackage[T1]{fontenc}
\usepackage[latin1]{inputenc}
\usepackage{graphicx}
\usepackage[english]{babel}
\usepackage{graphicx}
\usepackage{bm}
\usepackage{amsmath}
\usepackage{amssymb}
\usepackage{amsfonts}
\usepackage{epsfig}
\usepackage{colordvi}
\usepackage{color}

\def\be{\begin{equation}}
\def\ee{\end{equation}}
\def\bea{\begin{eqnarray}}
\def\eea{\end{eqnarray}}
\def\beq{\begin{eqnarray}}
\def\eeq{\end{eqnarray}}

\begin{document}
\title{\textbf{Jeans analysis of self-gravitating systems in $f(R)$ gravity}}
\author{S. Capozziello$^{1,2}$\footnote{e-mail address: capozziello@na.infn.it},
M. De Laurentis$^{1,2}$\footnote{e-mail address: mariafelicia.delaurentis@na.infn.it},
I. De Martino$^{3}$\footnote{e-mail address:ivan.demartino1983@gmail.com},
M. Formisano$^{4}$\footnote{e-mail address:mikformi@hotmail.com},
S.D. Odintsov$^5$\footnote{e-mail address: odintsov@aliga.ieec.uab.es} }
\affiliation{$^{1}$ Dipartimento di Scienze Fisiche, Universit\`a di Napoli "Federico II"}
\affiliation{$^{2}$ INFN sez. di Napoli Compl. Univ. di Monte S. Angelo, Edificio G, Via Cinthia, I-80126 - Napoli, Italy}
\affiliation{$^{3}$\it Departamento de  Fisica Teorica, University of Salamanca, 37008 Salamanca, Spain}
\affiliation{$^{4}$ \it Dipartimento di Fisica, Universit\`a di Roma "La Sapienza", Piazzale Aldo
Moro 5, I-00185 Roma, Italy}
\affiliation{$^{5}$Institucio Catalana de Recerca i Estudis Avancats (ICREA) and Institut de Ciencies de l Espai (IEEC-CSIC),
Campus UAB, Facultat de Ciencies, Torre C5-Par-2a pl, E-08193 Bellaterra (Barcelona), Spain}
\date{\today}

\begin{abstract}
Dynamics and collapse of collisionless self-gravitating  systems  is described by the coupled collisionless Boltzmann and Poisson
equations derived from $f(R)$ gravity in the weak field approximation. Specifically, we describe a system at equilibrium
by a time-independent distribution function $f_0(x,v)$ and two potentials $\Phi_0(x)$ and $\Psi_0(x)$
solutions of the modified Poisson and collisionless Boltzmann equations. Considering a small perturbation from the  equilibrium and
linearizing the field equations, it can be obtained a dispersion relation. A dispersion equation is achieved for neutral dust-particle
systems  where a generalized Jeans wave-number is obtained.  This analysis gives rise to unstable modes not present
in the standard Jeans analysis (derived assuming Newtonian gravity as weak filed limit of  $f(R)=R$). In this perspective, we discuss several self-gravitating
astrophysical systems whose dynamics could be fully addressed in the framework of  $f(R)$-gravity.
\end{abstract}

\maketitle

%%%%%%%%%%%%%%%%%%%%%%%%%%%%%%%%%%%%%%%%%%%%%%%%%%%
\section{Introduction}
\label{uno}
%%%%%%%%%%%%%%%%%%%%%%%%%%%%%%%%%%%%%%%%%%%%%%%%%%%

One of the fundamental goals of  modern cosmology is to probe  Einstein's General Relativity (GR) at any scale beyond the classical 
tests that confirmed such a theory in the weak field limit and at Solar System level.
GR is assumed as the standard  theory of gravity  describing   astrophysical structures up to the whole observed Universe;
however there are some inconsistencies at ultraviolet  scales (e.g. the  initial singularity, the quantum gravity issue) and infrared scales (e.g.
cosmic acceleration,  concordance problem, flatness problem, galaxy rotation curves, large scale structure, massive stars formation) 
that strongly suggest that Einstein's approach should be revised or at least extended. Furthermore, astrophysical observations of the last decades suggest
that new (dark)  ingredients are necessary to achieve a self-consistent cosmological model. In particular, the
observations suggest the Hubble flow is currently accelerating, and the simplest way to explain the cosmic acceleration
is to insert a cosmological constant ($\Lambda$) in the Friedmann-Robertson-Walker Cosmology \cite{Padmanabhan,Peebles,Sahni},
representing about $70\%$ of the total amount of energy. On the other hand, the galaxy rotation curves and the large scale structure
could be dynamically addressed by introducing huge amounts  of dark matter (about the 25$\%$ of the total matter). Only  5$\%$ of the cosmic budget is constituted by  standard matter
as stars, neutrinos, radiation, heavy elements and free cosmological hydrogen ad helium.
Alternative approaches to GR could be pursued with the aim to  explain  the observed acceleration and missing matter without introducing
new ingredients  up to now not observed at fundamental scales. The  so-called $f(R)$-gravity is considered as a possible,
straightforward mechanism to explain the cosmic acceleration without inserting  unknown elements as dark energy and dark matter, but extending
the geometric part of the field equations by relaxing the strict hypothesis that the gravitational action has to be restricted to $f(R)=R$ as in the Hilbert-Einstein one 
\cite{physrep,Capozziello2,felix1,Nojiri, review, Capozziello10}.

These theories have been investigated both at cosmological
scales and in the weak field limit \cite{Capozziello3,f(R)test,Cognola1, Cognola2,f(R)test2}.
It has been shown that a late accelerating behavior can be easily recovered \cite{saffari} and it can be related to an early inflationary
expansion \cite{Nojiri2}. Furthermore,  modifying the gravity action by assuming   non-linear Lagrangians, one obtain corrections to the
gravitational potential which can be useful for astrophysical phenomenology  at galactic scales. In particular,
without the introduction of dark matter, the rotation curves of spiral galaxies and the haloes of galactic clusters  can be
dynamically  addressed \cite{mnras,salzano,Boehmer, Mendoza, Sobouti}. Several of these extended models  reproduce Solar System
tests so they are not in conflict with GR  experimental results  but simply extend them \cite{Capozziello8,Capozziello9, pogosian}.

It is important to stress that $f(R)$ gravity  has interesting applications also in stellar astrophysics and could contribute to solve
several puzzles related to observed peculiar objects (e.g. magnetars, stars in the instability strips, protostars, etc. \cite{Cooney, Hu}),
structure and star formation \cite{poly, Chang}.

Here we analyze the Jeans instability for self-gravitating systems in $f(R)$-gravity coupled with perfect-fluid matter. The aim is to show that several
self-gravitating systems, in particular those involved in star formation (e.g. large molecular clouds or Bok globules), can be exactly
addressed in this framework by considering the corrections  to the Newtonian potential coming out from $f(R)$-gravity. This fact could
constitute a remarkable signature to retain or rule out these theories at astrophysical level.

 The paper is organized as follows. In Section \ref{due} the classical theory  of gravitational collapse
for dust-dominated systems is summarized.  In Section \ref{tre}, we  discuss the weak field   limit of $f(R)$-gravity obtaining corrections to the standard
Newtonian potential that can be figured out as two Newtonian potentials contributing to the dynamics. In Section \ref{quattro} we
recover the dispersion relation and Jeans mass limit while, in  Section \ref{cinque},  some self-gravitating dust  system are considered in this approach.
The difference between GR and $f(R)$-gravity are put in evidence, in particular the Jeans mass profiles with respect to the temperature.
We report a catalogue of observed molecular clouds in order to compare the classical Jeans mass  with the  $f(R)$ one. Finally, in Section \ref{sei}, results are discussed.

%%%%%%%%%%%%%%%%%%%%%%%%%%%%
%\section{Self-gravitational neutral grain system}
\section{Dust-dominated self-gravitating systems}
\label{due}
%%%%%%%%%%%%%%%%%%%%%%%%%%%%%%%%%%%%%
The  collapse of self-gravitational collisionless systems can be dealt with the  introduction  of coupled
collisionless Boltzmann and Poisson equations  (for details, see \cite{BT}):

\begin{equation}\label{Boltz_poisson1}
{\begin{array}{l}
\dfrac{{\partial f(\vec r,\vec v,t)}}{{\partial t}} + \left( {\vec v\cdot\vec \nabla _r } \right)f(\vec r,\vec v,t) - \qquad\qquad\qquad\\
\\
\qquad\qquad\qquad\qquad -\left( {\vec \nabla \Phi \cdot\vec \nabla _v } \right)f(\vec r,\vec v,t) = 0
\end{array}}
\end{equation}

\begin{equation}\label{Boltz_poisson2}
 \vec \nabla ^2 \Phi (\vec r,t) = 4\pi G\int {f(\vec r,\vec v,t)} d\vec v\,,
\end{equation}
where $\vec v$ and $\vec r$ mean 3-dimensional vectors in the spatial manifold.

A self-gravitating system at equilibrium is described by a time-independent distribution function $f_0(x,v)$ and a potential $\Phi_0(x)$ that are solutions
 of  Eq.\eqref{Boltz_poisson1} and \eqref{Boltz_poisson2}. Considering a small perturbation to this equilibrium:

\begin{eqnarray}
 &&  f(\vec r,\vec v,t) = f_0 (\vec r,\vec v) + \epsilon f_1 (\vec r,\vec v,t),
 \\
 &&  \Phi (\vec r,t) = \Phi _0 (\vec r) + \epsilon \Phi _1 (\vec r,t),
\end{eqnarray}

where $\epsilon \ll 1$ and by substituting in Eq. \eqref{Boltz_poisson1} and \eqref{Boltz_poisson2} and by linearizing, one obtains:
\begin{equation}\label{Linear_Boltz_poisson1}
\begin{array}{l}
 \dfrac{{\partial f_1 (\vec r,\vec v,t)}}{{\partial t}} + \vec v\cdot\dfrac{{\partial f_1 (\vec r,\vec v,t)}}{{\partial \vec r}} +  \\
  \\
-\vec \nabla \Phi _1 (\vec r,t)\cdot\dfrac{{\partial f_0 (\vec r,\vec v)}}{{\partial \vec v}} - \vec\nabla \Phi _0 (\vec r)\cdot\dfrac{{\partial f_1 (\vec r,\vec v,t)}}{{\partial \vec v}} = 0\,,
 \end{array}
\end{equation}
\begin{equation}\label{Linear_Boltz_poisson2}
 \vec \nabla ^2 \Phi _1 (\vec r,t) = 4\pi G\int {f_1 (\vec r,\vec v,t)} d\vec v\,,
\end{equation}

Since the equilibrium state is assumed to be homogeneous and time-independent, one can  set $f_0(\vec x, \vec v,t) = f_0(\vec v)$,
and  the so-called Jeans "swindle" to set $\Phi_0 = 0$. In Fourier components, Eqs.\eqref{Linear_Boltz_poisson1} and
\eqref{Linear_Boltz_poisson2} become:
\begin{eqnarray}
 &&- i\omega f_1  + \vec v\cdot\left( {i\vec kf_1 } \right) - \left( {i\vec k\Phi _1 } \right)\cdot\frac{{\partial f_0 }}{{\partial \vec v}} = 0\,, \\
 &&- k^2 \Phi _1  = 4\pi G\int {f_1 } d\vec v.
 \end{eqnarray}
By combining these equations, the dispersion relation
\begin{equation}\label{eqDISP}
{1 + \frac{{4\pi G}}{{k^2 }}\int {\dfrac{{\vec k\cdot\dfrac{{\partial f_0 }}{{\partial \vec v}}}}{{\vec v\cdot\vec k - \omega }}} d\vec v}=0;
\end{equation}
is obtained.
In the case of stellar systems, by assuming a Maxwellian distribution function for $f_0$, we have
\begin{equation}\label{Maxwellian}
f_0  = \frac{{\rho _0 }}{{(2\pi \sigma ^2 )^{\frac{3}{2}} }}e^{ - \dfrac{{v^2 }}{{2\sigma ^2 }}},
\end{equation}
 imposing that $\vec k=(k,0,0)$ and  substituting in  Eq.\eqref{eqDISP}, one gets:
\begin{equation}
1 - \frac{{2\sqrt {2\pi } G\rho _0 }}{{k\sigma ^3 }}\int {\dfrac{{v_x e^{ - \dfrac{{v_x^2 }}{{2\sigma ^2 }}} }}{{kv_x  - \omega }}} dv_x  = 0.
\end{equation}
By setting $\omega=0$,  the limit for instability is obtained:
\begin{equation}
     k^2(\omega=0)  = \dfrac{{4\pi G\rho _0 }}{{\sigma ^2 }} = k_J^2,
\end{equation}
by which it is possible to  define the Jeans mass ($M_J$) as the mass originally contained within a sphere of diameter $\lambda_J$:
\begin{equation}\label{MJ}
 M_J= \frac{4 \pi}{3} \rho_0 \left(\frac{1}{2} \lambda_J \right)^3,
\end{equation}
where
\begin{equation}
   \lambda_J^2= \dfrac{\pi \sigma^2}{G \rho_0}
   \label{length}
\end{equation}
is the Jeans length.
Substituting Eq. \eqref{length} into  Eq. \eqref{MJ}, we recover
\begin{equation}\label{Mass}
{{{ M}_J} = \dfrac{\pi }{6}\sqrt {\dfrac{1}{{{\rho _0}}}{{\left( {\dfrac{{\pi {\sigma ^2}}}{{G}}} \right)}^3}} }\,.
\end{equation}

All perturbations with wavelengths $\lambda >\lambda_J$  are unstable in the stellar system.  In order to evaluate the integral
in the dispersion relation, we have to study the singolarity at $\omega=kv_x$. To this end, it is useful to write 
the dispersion relation as
\begin{equation}
1 - \dfrac{{k_{_J }^2 }}{{k^2 }}W\left( \beta \right) = 0,
\end{equation}
defining
\begin{equation}
 W\left( \beta \right) \equiv \dfrac{1}{{\sqrt {2\pi } }}\int {\dfrac{{xe^{ - \frac{{x^2 }}{2}} }}{{x - \beta}}} dx,
\end{equation}
where $\beta= {\dfrac{\omega}{{k\sigma }}} $ and $x=\dfrac{v_x}{\sigma} $. We set also $\omega=i\omega_I$ 
and $Re\left[ W\left( {\dfrac{\omega}{{k\sigma }}} \right)\right]=0$, because we are interested in the unstable modes. 
These modes appear when the imaginary part of $\omega$ is greater than zero and in this case the integral
in the dispersion relation can be resolved just with previous prescriptions.

In order to study unstable modes (for  details, see Appendix B in \cite{BT}) we replace the following identities
\[
\left\{ \begin{array}{l}
 \int\limits_0^\infty  {\dfrac{{x^2 e^{ - x^2 } }}{{x^2  + \beta ^2 }}dx}  = \dfrac{1}{2}\sqrt \pi   - \dfrac{1}{2}\pi \beta e^{\beta ^2 } \left[ {1 - {\rm{erf }}\beta } \right]\,, \\
  \\
 {\rm{erf }}z = \dfrac{2}{{\sqrt \pi  }}\int\limits_0^z {e^{ - t^2 } dt}.  \\
 \end{array} \right.
\]
into the dispersion relation obtaining:
\begin{equation}
\small{k^2  = k_J^2 \left\{ {1 -  \dfrac{{\sqrt \pi \omega _I }}{{\sqrt 2 k\sigma }}e^{\left( {\dfrac{{\omega _I }}{{\sqrt 2 k\sigma }}} \right)^2 } \left[ {1 - {\rm{erf }}\left( {\frac{{\omega _I }}{{\sqrt 2 k\sigma }}} \right)} \right]} \right\}}.
\end{equation}
This is the standard dispersion relation describing the  criterion to collapse for infinite homogeneous fluid and stellar systems \cite{BT}.

%%%%%%%%%%%%%%%%%%%%%%%%%%%%%%%%%%%%%%%%%%%%%%%%%%%
\section{Newtonian limit of $f(R)$-gravity}
\label{tre}
%%%%%%%%%%%%%%%%%%%%%%%%%%%%%%%%%%%%%%%%%%%%%%%%%%%

As discussed in the Introduction,  $f(R)$-gravity is a straightforward   extension of GR by which it is possible, in principle, to recover good results of
GR without imposing  {\it a priori} the form of  gravitational Lagrangian, chosen to be $f(R)=R$ by Hilbert and Einstein. This means that
we do not impose a priori the gravitational  action but  it can be, in principle, re-constructed by generic curvature invariants and then matched with observations
(the simplest choice in this sense is to take into account an analytic  function  of the Ricci scalar $R$  \cite{review}).
However, from a genuine mathematical viewpoint,  the initial value problem of such theories has to be carefully addressed  in order
to achieve self-consistent results (see for example \cite{Capozziello10}).

Let us start with a general class of higher order theories given by the action
\begin{eqnarray}\label{FOGaction}
\mathcal{A}\,=\,\int d^{4}x\sqrt{-g}[f(R)+\mathcal{X}\mathcal{L}_m]\,,
\end{eqnarray}
where $f(R)$ is an analytic function of curvature invariant $R$  and $\chi=\frac{8\pi G}{c^4}$ is the usual coupling of gravitational field equations \cite{review}.
The term $\mathcal{L}_m$ is the minimally coupled ordinary matter
contribution. In the metric approach, the field equations are
obtained by varying (\ref{FOGaction}) with respect to
$g_{\mu\nu}$. We get:

\begin{equation}\label{fieldequationFOG}
{\begin{array}{l}
f'(R)R_{\mu\nu}-\frac{1}{2}f(R)g_{\mu\nu}-\nabla_{\mu}\nabla_{\nu} f'(R)+\qquad\qquad\qquad\\\\
\qquad\qquad\qquad\qquad\qquad+g_{\mu\nu}\Box f'(R)=\mathcal{X}\,T_{\mu\nu}\,,
\end{array}}
\end{equation}
with the trace equation
\begin{eqnarray}
3\Box
f'(R)+f'(R)R-2f(R)\,=\,\mathcal{X}\,T.
\label{TRACE}
\end{eqnarray}
Here,
$T_{\mu\nu}\,=\,\frac{-1}{\sqrt{-g}}\frac{\delta(\sqrt{-g}\mathcal{L}_m)}{\delta
g^{\mu\nu}}$ is the the energy-momentum tensor of matter, while
$T\,=\,T^{\sigma}_{\,\,\,\,\,\sigma}$ is the trace,
$\Box\,=\,{{}_{;\sigma}}^{;\sigma}$ and $f'(R)\,=\,\frac{df(R)}{dR}$\footnote{Here we shall adopt  the convention
$c\,=\,1$. The convention for Ricci's tensor is
$R_{\mu\nu}\,=\,{R^\sigma}_{\mu\sigma\nu}$ while for the Riemann
tensor is
${R^\alpha}_{\beta\mu\nu}\,=\,\Gamma^\alpha_{\beta\nu,\mu}+...$. The
affinities are the usual Christoffel symbols of the metric
$\Gamma^\mu_{\alpha\beta}\,=\,\frac{1}{2}g^{\mu\sigma}(g_{\alpha\sigma,\beta}+g_{\beta\sigma,\alpha}
-g_{\alpha\beta,\sigma})$.}. The  signature is $(-\,+\,+\,+)$  \cite{landau} ).
For our purposes,    we have to start by setting the right approximation
in the metric tensor $g_{\mu\nu}$  \cite{gravitation}:

\begin{eqnarray}\label{metric_tensor_PPN}
  g_{\mu\nu}\,\sim\,\begin{pmatrix}
  -(1+2\,\Phi(t,\mathbf{x}))+\mathcal{O}(4)& \mathcal{O}(3) \\
  \\
  \mathcal{O}(3)& \delta_{ij}+\mathcal{O}(2)\end{pmatrix},
\end{eqnarray}
where $\mathcal{O}(n)$ (with $n\,=\,$integer) denotes the order of the expansion.
It is worth stressing that the expansion  parameter  is $c^{-1}$ and, in the Newtonian limit, we are assuming perturbations up to $c^{-2}$. This means that in the above expression (\ref{metric_tensor_PPN}), we can discard terms of order $\mathcal{O}(3)$ and $\mathcal{O}(4)$ that have to be considered in further perturbation post-Newtonian limit (see \cite{physrep} and references therein).
The set of coordinates\footnote{The Greek index runs between $0$
and $3$; the Latin index between $1$ and $3$.} adopted is
$x^\mu\,=\,(t,x^1,x^2,x^3)$. The Ricci scalar becomes

\begin{eqnarray}
R\,\sim\,R^{(2)}(t,\mathbf{x})+\mathcal{O}(4)\,.
\end{eqnarray}
The $n^{th}$ derivative of Ricci function can be developed as

\begin{eqnarray}
f^{n}(R)\,&\sim&\,f^{n}(R^{(2)}+\mathcal{O}(4))\,\sim\nonumber\\&&
\sim\,f^{n}(0)+f^{n+1}(0)R^{(2)}+\mathcal{O}(4)\,,\nonumber\\
\end{eqnarray}
here $R^{(n)}$ denotes a quantity of order  $\mathcal{O}(n)$.
It is worth stressing that the symbol $f^n(R)$ means the $n^{th}$ derivative of the analytic function $f(R)$. In the following, we are going to use the numbers $f^n(0)$ of the Taylor series.
From lowest order of field equations (\ref{fieldequationFOG}), we
have $f(0)\,=\,0$ which trivially follows from the above assumption (\ref{metric_tensor_PPN}) that the space-time
is asymptotically Minkowskian. Eqs.(\ref{fieldequationFOG})  and (\ref{TRACE}) at $\mathcal{O}(2)$-order (Newtonian level)
become:

\begin{eqnarray}\label{PPN-field-equation-general-theory-fR-O2}
&&R^{(2)}_{tt}-\frac{R^{(2)}}{2}-f''(0)\nabla^2
R^{(2)}\,=\,\mathcal{X}\,T^{(0)}_{tt}\,,
\\
&&-3f''(0)\nabla^2
R^{(2)}-R^{(2)}\,=\,\mathcal{X}\,T^{(0)}\,,
\end{eqnarray}
where $\nabla$ is the Laplacian in the flat space, $R^{(2)}_{tt}\,=\,\nabla^2\Phi(t,\mathbf{x})$ and for the sake of simplicity,  we set $f'(0)\,=\,1$. 
We recall that the energy-momentum  tensor for a perfect fluid  is the 

\begin{eqnarray}
T_{\mu\nu}\,=\,(\epsilon+p)\,u_{\mu} u_{\nu}-p\,g_{\mu\nu}\,,
\end{eqnarray}
where $p$ is the pressure  and $\epsilon$ is the energy density. If we consider a perfect fluid of dust ($p=0$), we have $R^{(2)}_{00}=\frac{1}{2}\nabla^2 g_{00}$ 
\cite{gravitation}. Then we have

\begin{eqnarray}\label{HOEQ}
&&\nabla^2 \Phi-\frac{R^{(2)}}{2}-f''(0)\nabla^2 R^{(2)}\,= \,\mathcal{X}\rho\,,
\\
&&-3f''(0)\nabla^2 R^{(2)}-R^{(2)}\,=\,\mathcal{X}\rho\,,\label{HOEQTRA}
\end{eqnarray}
where $\rho$ is the mass density\footnote{We remember that  $\epsilon\,=\,\rho\,c^2$.}. For $f''(0)\,=\,0$, the standard  Poisson equation $\nabla^2\Phi\,=\,4\pi G\rho$ is recovered.

The solution for the gravitational potential $\Phi$ has
a Yukawa-like behavior depending on a
characteristic length. Then as it is evident,
the Gauss theorem is not valid since the force
law is not $\propto|\mathbf{x}|^{-2}$. The equivalence between a
spherically symmetric distribution and point-like distribution is
not valid and how the matter is distributed in the space is very
important\footnote{However, we have to see that being the Yukawa correction a decreasing exponential function, the Gauss theorem is asymptotically recovered. 
In any case, conservation laws are always preserved since the Bianchi identities hold.}.

Besides the Birkhoff Theorem at Newtonian level is modified:
the solution can be only factorized with a  space-depending function
and an arbitrary  time-depending function. Furthermore,
the correction to the gravitational potential is depending on
the first two derivatives of $f(R)$ in $R\,=\,0$. So different analytical 
models,  up to the third derivative, admit the same Newtonian general
solution.

Field equations (\ref{HOEQ}) and (\ref{HOEQTRA}) give rise to the \emph{modified Poisson equations}
for $f(R)$-gravity.
%Now for our purpose we restore the factor $c$ and then the (\ref{HOEQ}) becomes:
%\begin{eqnarray}\label{HOEQ1}
%&&\frac{1}{c^2}\nabla^2 \Phi-\frac{R^{(2)}}{2}-f''(0)\nabla^2 R^{(2)}\,= \,\mathcal{X}\rho c^2\,,
%\\
%&& -3f''(0)\nabla^2 R^{(2)}-R^{(2)}\,=\,\mathcal{X}\rho c^2\,,
%\end{eqnarray}
%with simple steps we get from the second equation $\nabla R^{(2)}$
%%\begin{equation}
%%f''(0)\nabla^2 R^{(2)}=-\frac{\mathcal{X}\rho c^2+R^{(2)}}{3}\,,
%%\end{equation}
%and replace it in the first equation to obtain
%\begin{equation}
%\nabla^2 \Phi-\frac{c^2 R^{(2)}}{6}=\frac{16}{3}\pi G \rho\,.
%\end{equation}
%After some calculations we obtain:
We know that
\begin{eqnarray}
R^{(2)}&\simeq \frac{1}{2}\nabla^2 g^{(2)}_{00}- \frac{1}{2}\nabla^2g^{(2)}_{ii}\,.\label{3}
\end{eqnarray}
Inserting in the above result the $g_{\mu\nu}$ approximations \eqref{metric_tensor_PPN} we obtain
\begin{eqnarray}
{R^{(2)}} \simeq {\nabla ^2}(\Phi  - \Psi )\,,
\end{eqnarray}
where $\Psi$ is the further gravitational potential related to the metric component $g^{(2)}_{ii}$.
%Finally we can write the Poisson equation
%\begin{eqnarray}\label{NewPoisson}
%\nabla^2 \Phi=8\pi G\rho\,.
%\end{eqnarray}
%Recalling that $R^{(2)}=-2\nabla^2 \Phi$ (see \cite{landau})and
 Substituting in  Eqs. (\ref{HOEQ}) and (\ref{HOEQTRA}), we obtain
\begin{eqnarray}\label{EQP}
&& \nabla^2\Phi+\nabla^2\Psi -2 f''(0) \nabla^4 \Phi + 2 f''(0) \nabla^4 \Psi=2\mathcal{X} \rho \\
\nonumber \\
&& \nabla^2\Phi-\nabla^2\Psi +3 f''(0)\nabla^4\Phi-3 f''(0) \nabla^4 \Psi=-\mathcal{X} \rho \label{EQP2}\,.
\end{eqnarray}
By eliminating the higher-order terms,  the standard Poisson equation is recovered.
Our  task is to check how the Jeans instability occurs in $f(R)$-gravity.

An important consideration is in order at this point.  As we pointed out above, we are supposing that the space-time is asymptotically Minkowski.
However, this is
against the general idea of $f(R)$-gravity which should   mimic  dark energy behavior. This means that the space-time should be
asymptotically de Sitter. So, in general, it is necessary that the
$f(R)$ function is expandable at $R=0$, or even if it is, the interesting asymptotic
space is nevertheless not Minkowskian ($R=0$) but $R\ne 0$. This fact is 
connected with the
assumption that the energy-density $\rho$ is homogenous and asymptotically constant in order to leads
to de Sitter space-time. This is, at the very end, why the so called ``Jeans swindle'' is
needed in Newtonian theory. In the present case,  $\rho$ is explicitly written in Eqs. (\ref{HOEQ}) and
(\ref{HOEQTRA}) and has to converge asymptotically to zero in order to restore the asymptotic Minkowskian behavior. 
 In other words,  the possible gravitational
actions have to be chosen so that the condition $f(0)=0$ holds.

%%%%%%%%%%%%%%%%%%%%%%%%%%%%%%%%%%%%%%%%%%%%%%%%%%%
\section{Jeans criterion for gravitational instability in $f(R)$ gravity}
\label{quattro}
%%%%%%%%%%%%%%%%%%%%%%%%%%%%%%%%%%%%%%%%%%%%%%%%%%%

\begin{figure}[!h]
 \centering
  \includegraphics[scale=0.39]{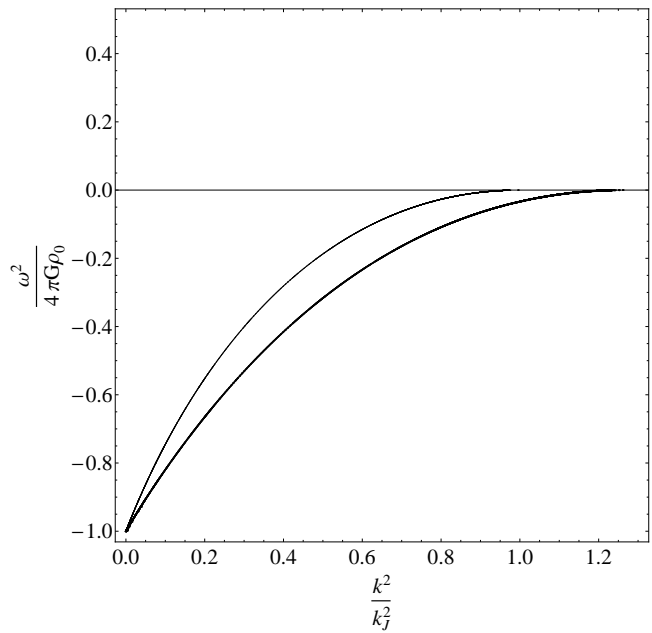}\\
  \caption{The bold line indicates the plot of the dispersion relation \eqref{eq:Results} in which we imposed
the value for $\alpha$ given by \eqref{alpha}. The thin line indicates the plot of the standard dispersion equation
\cite{BT}.}\label{Fig:Jeans}
\end{figure}

Our task is now to study the Jeans instability in the framework of $f(R)$ gravity.
 Let us  assume the standard collisionless Boltzmann equation:
\begin{equation}
{\begin{array}{l}
\dfrac{{\partial f(\vec r,\vec v,t)}}{{\partial t}} + \left( {\vec v\cdot\vec \nabla _r } \right)f(\vec r,\vec v,t) + \qquad\qquad\qquad\\
\\
\qquad\qquad\qquad\qquad -\left( {\vec \nabla \Phi \cdot\vec \nabla _v } \right)f(\vec r,\vec v,t) = 0\,,
\end{array}}
\end{equation}
where, according to the Newtonian theory, only the potential $\Phi$ is present. 
Considering the  $f(R)$ Poisson equations, given by  Eqs. \eqref{EQP} and \eqref{EQP2}, also the potential $\Psi$ 
has to be considered so we obtain the coupled equations

\begin{equation}
 \nabla^2(\Phi+\Psi) -2 \alpha \nabla^4 (\Phi -\Psi)=16\pi G\int {{f}(\vec r,\vec v,t)} d\vec v{\mkern 1mu}
\end{equation}
\begin{equation}
 \nabla^2(\Phi-\Psi) +3 \alpha \nabla^4(\Phi-\Psi)=-8\pi G\int {{f}(\vec r,\vec v,t)} d\vec v{\mkern 1mu}\,.
\end{equation}
In the previous equations, we have replaced  $f''(0)$ with the greek letter $\alpha$. It is important to stress that while in the standard theory, the distribution function $f(\vec r,\vec v,t)$ is related only to the potential $\Phi$, it is related to both $\Phi$ and $\Psi$ in the Newtonian limit coming from $f(R)$ gravity.
As in standard case, we consider small perturbation to the equilibrium and linearize the equations. After
we write equations  in Fourier space so they became
\begin{eqnarray}\label{F1}
&&{ - i\omega {f_1} + \vec v\cdot\left( {i\vec k{f_1}} \right) - \left( {i\vec k{\Phi _1}} \right)\cdot\frac{{\partial {f_0}}}{{\partial \vec v}} = 0},
\\ \label{F2}
&&{  - {k^2}{(\Phi _1} +{\Psi _1}) - 2\alpha {k^4}({\Phi _1} -{\Psi _1}) = 16\pi G\int {{f_1}} d\vec v},
\\ \label{F3}
&& {  {k^2}({\Phi _1} -{\Psi _1}) - 3\alpha {k^4}({\Phi _1} -{\Psi _1}) =  8\pi G\int {{f_1}} d\vec v}.
\end{eqnarray}
Combining Eqs. \eqref{F2} and\eqref{F3}, we obtain a relation between $\Phi_1$ and $\Psi_1$,
\begin{equation*}
{{\Psi _1} =  \frac{{3 - 4\alpha {k^2}}}{{1 - 4\alpha {k^2}}}{\Phi _1}}
\end{equation*}
inserting this relation in Eq. \eqref{F2} and combining it with Eq. \eqref{F1}, we obtain the dispersion relation
\begin{equation}\label{newEQdispersion}
1 - 4\pi G\dfrac{{1 - 4\alpha {k^2}}}{{3\alpha {k^4} - {k^2}}}\int {\left( {\dfrac{{\vec k\cdot\dfrac{{\partial {f_0}}}{{\partial \vec v}}}}{{\vec v\cdot\vec k - \omega }}} \right)} d\vec v = 0.
\end{equation}
If we assume, as in standard case, that $f_0$ is given by \eqref{Maxwellian} and $\vec{k} = (k,0,0)$, one can write
\begin{equation}\label{NewEqDispersion2}
{1 + \frac{{2\sqrt {2\pi } G{\rho _0}}}{{{\sigma ^3}}}\dfrac{{1 - 4\alpha {k^2}}}{{3\alpha {k^4} - {k^2}}}\left[ {\int {\dfrac{{k{v_x}{e^{ - \frac{{v_x^2}}{{2{\sigma ^2}}}}}}}{{k{v_x} - \omega }}} d{v_x}} \right] = 0}.
\end{equation}
By eliminating the higher-order terms (imposing $\alpha=0$), we obtain again  the standard dispersion Eq. \eqref{eqDISP}.
In order to compute the integral in the dispersion relation \eqref{NewEqDispersion2}, we consider the same approach  used
in the classical case, and finally we obtain:
\begin{equation}\label{eq:Results}
{\begin{array}{l}
1 + \mathcal{G}{\dfrac{{1 - 4\alpha {k^2}}}{{3\alpha {k^4} - {k^2}}}} \left[ {1 - \sqrt\pi x e^{x^2}\left( {1 - {\rm{erf}}\left[ x\right]} \right)} \right] = 0,
\end{array}}
\end{equation}
where $x=\dfrac{{{\omega _I}}}{{\sqrt 2 k\sigma }}$ and $\mathcal{G}=\dfrac{{4G\pi {\rho _0}}}{{{\sigma ^2}}}$.
In order to  evaluate Eq. \eqref{eq:Results} comparing it with the classical one, given by Eq. \eqref{eqDISP}, it is very useful to
normalize the equation to the classical Jeans length showed in Eq. \eqref{length}, by fixing the parameter of  $f(R)$-gravity, that is
\begin{equation}\label{alpha}
    \alpha   =  - \frac{1}{{k_j^2}}=  - \frac{{{\sigma ^2}}}{{4\pi G{\rho _0}}}.
\end{equation}
This parameterization is correct because the dimension $\alpha$ (an inverse of squared length) allows us to parametrize  as in standard case.
Finally we  write
\begin{equation}
\begin{array}{l}\label{instability}
\dfrac{{3{k^4}}}{{k_j^4}}+ \dfrac{{{k^2}}}{{k_j^2}} = \left( {\dfrac{{4{k^2}}}{{k_j^2}} + 1} \right) \left[ {1 - \sqrt\pi x e^{x^2}\left( {1 - {\rm{erf}}\left[ x\right]} \right)} \right] =0.
\end{array}
\end{equation}
The function is plotted in Fig.\ref{Fig:Jeans}, where Eq. \eqref{eq:Results} and the
standard dispersion  \cite{BT} are confronted in order to see the difference between $f(R)$ and Newtonian gravity.

As shown in  Fig. \ref{Fig:Jeans}, the effects of a different theory of gravity changes the limit of instability. The limit is higher than
the classical case and the curve has a greater slope. This fact is important because  the mass limit value of
 interstellar clouds decreases changing the initial conditions to start the collapse.

%%%%%%%%%%%%%%%%%%%%%%%%%%%%%%%%%%%%%%%%%%%%%%%%%%%%%%%%%%%%
\section{The Jeans mass limit in $f(R)$-gravity}
\label{cinque}
%\section{Estimating the Jeans Mass limit in f(r)-gravity}
%%%%%%%%%%%%%%%%%%%%%%%%%%%%%%%%%%%%%%%%%%%%%%%%%%%%%%%%%%%%

 A numerical estimation of the $f(R)$-instability length in terms of the standard Newtonian one can be achieved. By solving numerically  Eq. \eqref{instability} with the condition $\omega=0$, we obtain that the collapse occurs for
\begin{equation}\label{newlengthnumerical}
    k^2=1.2637 k^2_J.
\end{equation}
However we can estimate also analytically the limit for the instability. In order to evaluate the Jeans mass limit in $f(R)$-gravity, we set $\omega=0$ in  Eq. \eqref{NewEqDispersion2} and then
\begin{equation}\label{NewMass}
3{\sigma ^2}\alpha {k^4} - \left( {16\pi G{\rho _0}\alpha  + {\sigma ^2}} \right){k^2} + 4\pi G{\rho _0} = 0.
\end{equation}
It is worth stressing that the additional condition $\alpha<0$   discriminates  the class of viable $f(R)$ models: in such a case   we obtain stable cosmological solution
and positively defined massive states \cite{Capozziello10}.  In other words, this condition selects  the physically viable models allowing  to
solve Eq.\eqref{NewMass} for real values of k. In particular, the above numerical solution can be recast as 
\begin{equation}
    k^2=\frac{2}{3} \left(3+\sqrt{21}\right) \pi  \frac{G \rho }{\sigma ^2}.
\end{equation}
The relation to the Newtonian value of the Jeans instability is
\begin{equation}
    k^2=\frac{1}{6} \left(3+\sqrt{21}\right) k^2_J.
\end{equation}
Now, we can define the new Jeans mass as:
\begin{equation}\label{newMass2}
    \tilde{M}_J=6\sqrt{\frac{6}{\left(3+\sqrt{21}\right)^3}}  M_J,
\end{equation}
that is proportional to the standard Newtonian value. We will confront this specific solutions with some observed structures.

%%%%%%%%%%%%%%%%%%%%%%%%%%%%%%%%%%%%%%%%%%%%%%%%%%%
\subsection*{V.1 The  $M_J$ - $T$ relation}
\label{cinque.uno}
%%%%%%%%%%%%%%%%%%%%%%%%%%%%%%%%%%%%%%%%%%%%%%%%%%%
 Star formation is one of the best settled problems of modern astrophysics. However, some shortcomings emerge as soon as one faces  dynamics of  diffuse gas evolving  into stars and  star formation in  galactic environment.
One can deal with the star formation problem in two ways:
$i)$ we  can take into account the formation of  individual stars  and $ii)$ we can  discuss the formation of the  whole star  system   starting from interstellar clouds \cite{Mckee2}.
To answer these problems it is very important to study the interstellar medium (ISM) and its properties.The ISM physical conditions 
in the galaxies change in a very wide range, from hot X-ray emitting plasma to cold molecular gas, so it is very complicated to classify
the ISM by its properties. However, we can distinguish, in the first approximation, between \cite{carroll, kipp, dopita, Scheff}:
\begin{itemize}
\item {\bf Diffuse Hydrogen Clouds}. The most powerful tool to measure the properties of these clouds is the 21cm line emission of HI.
      They are cold clouds so the temperature is in the range $10\div 50$ K, and their  extension is up to  $50\div 100$ kpc from galactic center.
\item {\bf Diffuse Molecular Clouds}  are generally self-gravitating, magnetized, turbulent fluids systems, observed in sub-mm.  The most of the molecular gas is $H_2$, and the rest is CO. Here, the
      conditions are very similar to the HI clouds but in this case, the cloud can be more massive. They have, typically, masses in the range $3\div100 \,M_\odot$, temperature in $15\div 50$K and particle density in $(5\div50) \times 10^8$ m$^{-3}$.
\item {\bf Giant Molecular Clouds}  are very large complexes of particles (dust and gas), in which the range of the masses is typically $10^5\div 10^6\, M_\odot$ but they are very cold. The temperature is $\sim15$K, and the number of particles is $(1\div 3) \times 10^8$ m$^{-3}$ \cite{Mckee, Mckee2, Blitz, Blitz2}.
      However, there  exist also  small molecular clouds with masses $M < 10^4 M_\odot$\cite{Blitz2}.
They are the best sites for star formation, despite the mechanism of formation does not recover the star formation rate that  would be $250 M_\odot yr^{-1}$ \cite{Mckee}.
\item {\bf HII regions}. They are ISM regions with temperatures in the range  $10^3 \div 10^4$ K, emitting primarily in the radio and IR regions. At low frequencies,
       observations are associated to free-free electron transition (thermal Bremsstrahlung).
      Their densities range from over a million particles per cm$^3$ in the ultra-compact H II regions to only a few particles per cm$^3$
      in the largest and most extended regions. This implies total masses between $10^2$ and $10^5$ M$_\odot$ \cite{HII}.
\item {\bf Bok Globules}  are dark clouds of dense cosmic dust and gas in which star formation sometimes takes place. Bok globules are found within
      H II regions, and typically have a mass of about $2$ to $50$  M$_\odot$ contained within a region of about a light year.
\end{itemize}

\begin{figure}[!hl]
 \centering
  \includegraphics[width=1.05\columnwidth]{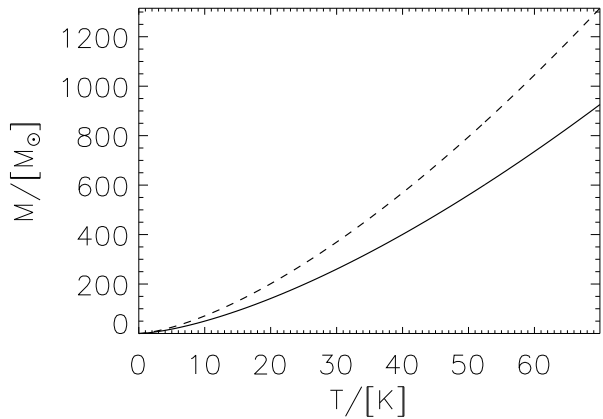}
  \caption{The  $M_J$-$T$ relation. Dashed-line indicates  the Newtonian Jeans mass behavior with respect to the temperature.
 Continue-line indicates the same for   $f(R)$-gravity Jeans mass.}\label{Fig:mass}
\end{figure}

Using  very general conditions \cite{carroll, kipp, dopita, Scheff, Mckee, Mckee2, Blitz, Blitz2,HII}, we want to show
the difference in the Jeans mass value between standard  and $f(R)$-gravity. Let us  take into account   Eq. \eqref{Mass} and  Eq.\eqref{newMass2}:
\begin{equation}\label{mass3}
{{{ M}_J} = \frac{\pi }{6}\sqrt {\frac{1}{{{\rho_0}}}{{\left( {\frac{{\pi {\sigma ^2}}}{{G}}} \right)}^3}} },
%\\
%&& {{{\tilde M}_J} = 6\sqrt{\frac{6}{\left(3+\sqrt{21}\right)^3}}  M_J},\label{mass4}
\end{equation}
in which $\rho_0$ is the ISM density and $\sigma$ is  the velocity dispersion of  particles due to the
temperature. These two quantities are defined as
$$\rho_0= m_{H} n_{H} \mu, \rm{\quad \quad} \sigma^2=\frac{k_BT}{m_H}$$
where $n_H$ is the number of particles measured in $m^{-3}$, $\mu$ is the mean molecular weight, $k_B$ is the Boltzmann constant and
$m_H$ is the proton mass. By using these relations, we are able to compute the Jeans mass for interstellar clouds and to plot its behavior 
 against  the temperature. Results are shown in Tab.\ref{table1} and Fig.\ref{Fig:mass}.  We have plotted the relation
for GR and for   $f(R)$ gravity. Any astrophysical system  reported  in Tab.\ref{table1} is associated to a particular 
 $(M,T)$-region. Differences between the two theories for any self-gravitating system are clear.

\begin{table}[!hl]
\small{\begin{tabular}[c]{|l|ccccc|}
\hline
Subject			  &       T   &         n          &   $\mu$    &       $M_J$     &  $\tilde{M}_J$ \\
		          &      (K)  &  ($10^8$m$^{-3}$)  &            &    ($M_\odot$)  &  ($M_\odot$)   \\
\hline
Diffuse Hydrogen Clouds   &      50   &   	5.0   	   &     1  	&      795.13     &    559.68     \\
Diffuse Molecular Clouds  &      30   &   	50   	   &     2 	&      82.63 	  &     58.16	   \\
Giant Molecular Clouds    &      15   &   	1.0   	   &     2 	&      206.58 	  &    145.41    \\
Bok Globules  		  &      10   &   	100 	   &     2 	&      11.24      &      7.91   \\
\hline
\end{tabular}}
\caption{ Jeans masses derived  from Eq. \eqref{Mass} (Newtonian gravity)  and \eqref{newMass2}  ($f(R)$-gravity).} \label{table1}
\end{table}

\begin{center}
\begin{table}[!h]
\small{\begin{tabular}[c]{|l|cccc|}
\hline
     Subject	  &       T    &         n     &     $M_J$     &  $\tilde{M}_J$ \\
		          &      K  &  ($10^8$m$^{-3}$)  &    ($M_\odot$)  &  ($M_\odot$)   \\
\hline
GRSMC G053.59+00.04 & 5.97 & 1.48  &  18.25 &  12.85 \\
GRSMC G049.49-00.41 & 6.48 & 1.54  &  21.32 &  15.00 \\
GRSMC G018.89-00.51 & 6.61 & 1.58  &  22.65 &  15.94 \\
GRSMC G030.49-00.36 & 7.05 & 1.66  &  22.81 &  16.06 \\
GRSMC G035.14-00.76 & 7.11 & 1.89  &  28.88 &  20.33 \\
GRSMC G034.24+00.14 & 7.15 & 2.04  &  29.61 &  20.84 \\
GRSMC G019.94-00.81 & 7.17 & 2.43  &  29.80 &  20.98 \\
GRSMC G038.94-00.46 & 7.35 & 2.61  &  31.27 &  22.01 \\
GRSMC G053.14+00.04 & 7.78 & 2.67  &  32.06 &  22.56 \\
GRSMC G022.44+00.34 & 7.83 & 2.79  &  32.78 &  23.08 \\
GRSMC G049.39-00.26 & 7.90 & 2.81  &  35.64 &  25.09 \\
GRSMC G019.39-00.01 & 7.99 & 2.87  &  35.84 &  25.23 \\
GRSMC G034.74-00.66 & 8.27 & 3.04  &  36.94 &  26.00 \\
GRSMC G023.04-00.41 & 8.28 & 3.06  &  38.22 &  26.90 \\
GRSMC G018.69-00.06 & 8.30 & 3.62  &  40.34 &  28.40 \\
GRSMC G023.24-00.36 & 8.57 & 3.75  &  41.10 &  28.93 \\
GRSMC G019.89-00.56 & 8.64 & 3.87  &  41.82 &  29.44 \\
GRSMC G022.04+00.19 & 8.69 & 4.41  &  47.02 &  33.10 \\
GRSMC G018.89-00.66 & 8.79 & 4.46  &  47.73 &  33.60 \\
GRSMC G023.34-00.21 & 8.87 & 4.99  &  48.98 &  34.48 \\
GRSMC G034.99+00.34 & 8.90 & 5.74  &  50.44 &  35.50 \\
GRSMC G029.64-00.61 & 8.90 & 6.14  &  55.41 &  39.00 \\
GRSMC G018.94-00.26 & 9.16 & 6.16  &  55.64 &  39.16 \\
GRSMC G024.94-00.16 & 9.17 & 6.93  &  56.81 &  39.99 \\
GRSMC G025.19-00.26 & 9.72 & 7.11  &  58.21 &  40.97 \\
GRSMC G019.84-00.41 & 9.97 & 11.3  &  58.52 &  41.19 \\
\hline
\end{tabular}}
\caption{We report the name,  the particle number density and the excitation temperature of observed molecular clouds. For each  system, we have calculated  the value of Jeans mass   in both  Newtonian (GR) and $f(R)$ gravity. The differences between the two approaches are significant pointing out that the star formation efficiency strictly depends on the adopted theory.
This table is only a part of the catalog of molecular clouds reported in  \cite{duval}.}\label{table2}
\end{table}
\end{center}
By using Eq.\eqref{newMass2} and by referring to the catalog of molecular clouds in Roman-Duval J. et al. \cite{duval}, we have
calculated the Jeans mass in the Newtonian and $f(R)$ cases. Tab.\ref{table2} shows the results. In all cases we note a substantial
difference between the classical and $f(R)$ value. In $f(R)$ scenario, molecular clouds become sites where star formation is strongly
supported and more efficient because in each of them the limit for the gravitational collapse is lower than the one in 
GR.

%\begin{widetext}
%
%\begin{figure}[!h]
%  \centering
%  \includegraphics[scale=1]{mass.eps}
%  \caption{We show the behavior of the relation $M_J$-$T$. Dot-line indicates the  Newtonian mass Jeans limit, the continue-line  the $f(R)$-gravity mass 
%  Jeans limit for the different systems considered.}\label{Fig:mass}
%\end{figure}
%
%\end{widetext}

%%%%%%%%%%%%%%%%%%%%%%%%%%%%%%%%%%%%%%%%%%%%%%%%%%%
\section{Discussion and Conclusions}
\label{sei}
%%%%%%%%%%%%%%%%%%%%%%%%%%%%%%%%%%%%%%%%%%%%%%%%%%%

 $f(R)$-gravity is an approach aimed to address some shortcomings of modern cosmology just assuming extensions of GR without invoking the presence of dark ingredients. 
 In other words, dark energy and dark matter could be effects related to curvature further degrees of freedom instead of new fundamental particles.

Here we have analyzed the Jeans instability mechanism, adopted for star formation,  considering the
Newtonian approximation of  $f(R)$-gravity. The related Boltzmann-Vlasov system leads to  modified
Poisson equations depending on the $f(R)$-model. In particular, 
considering  Eqs.\eqref{EQP} and
\eqref{EQP2},  it is possible to get a new dispersion relation \eqref{eq:Results} where 
instability criterion results modified (see also \cite{poly}). The leading parameter is $\alpha$, i.e.  the second derivative of the specific $f(R)$-model. Standard Newtonian Jeans instability is immediately recovered for $\alpha=0$ corresponding to the Hilbert-Einstein Lagrangian of GR.
 In Fig. \ref{Fig:Jeans}, dispersion relations for Newtonian and a specific $f(R)$-model are numerically compared. 
 The modified characteristic length van be given in terms of the classical one. 

Both in the classical  and  in $f(R)$ analysis,
the system damps the perturbation. This damping is not associated to the collisions
because we neglect them in our treatment, but it is linked to the so called Landau
damping \cite{BT}.

A new condition for the gravitational instability is derived, showing  unstable modes with faster
growth rates. Finally we can observe the instability decrease 
in $f(R)$-gravity: such decrease is related to a larger
Jeans length and then to a lower Jeans mass. We have also compared the
behavior with the temperature of the Jeans mass for various types of interstellar
molecular clouds (Fig. \ref{Fig:mass}). In 
Tables \ref{table1} and \ref{table2} we show the results given by this
new limit of the Jeans mass for a sample of giant molecular clouds. In
our model the limit (in unit of mass) to start the collapse of an
interstellar cloud is lower than the classical one advantaging the
structure formation.  Real solutions for the Jean mass can be achieved only for
$\alpha<0$  and this result is in agreement with cosmology \cite{Capozziello10}.
In particular,  the condition $\alpha<0$ is essentials to have
a  well-formulated and well-posed Cauchy problem in $f(R)$-gravity  \cite{Capozziello10}.
Finally, it is worth noticing that  the Newtonian value is an upper limit for the Jean mass coinciding with $f(R)=R$.

This work is intended to indicate the possibility to deal with ISM collapsing clouds  under
different assumptions about gravity. It is important to stress that we  fully recover the 
standard  collapse mechanisms but we could also describe proto-stellar systems that escape the standard collapse model. On the other hand,  this is the first step to study star formation in alternative theories of gravity (see also \cite{Cooney, Hu, poly, Chang,agon}). From an observational point of view, 
reliable  constraints can be achieved from  a careful analysis of the proto-stellar phase taking into 
account magnetic fields, turbulence and collisions. Finally, addressing  stellar systems by this approach could be an extremely important to test observationally $f(R)$-gravity.

Moreover, the approach developed in this work admits direct generalizations
for other modified gravities, like non-local gravity,
modified Gauss-Bonnet theory, string-inspired gravity, etc. In these cases,
the constrained Poisson equation may be even more complicated due to the
presence of extra scalar(s) in non-local or string-inspired gravity.
Developing further this approach gives, in general, the possibility to confront the
observable  dynamics of astrophysical objects (like stars) with predictions
of alternative gravities.

\end{document}